\documentclass[]{spie}  
\usepackage[]{graphicx}

\title{Characterization of an x-ray hybrid CMOS detector with low interpixel capacitive crosstalk  } 

\author{Christopher V. Griffith\supit{a}, Stephen D. Bongiorno\supit{a}, David N. Burrows\supit{a}, Abraham D. Falcone\supit{a}, Zachary R. Prieskorn\supit{a} 
\skiplinehalf
\supit{a}The Pennsylvania State University, 525 Davey Lab, University Park, PA, USA; \\
}

\authorinfo{Further author information: (Send correspondence to C.V.G.)\\C.V.G.: E-mail: cvg5124@psu.edu\\ }

\begin{document} 
  \maketitle 

\begin{abstract}
We present the results of x-ray measurements on a hybrid CMOS detector that uses a H2RG ROIC and a unique bonding structure. The silicon absorber array has a 36$\mu$m pixel size, and the readout array has a pitch of 18$\mu$m; but only one readout circuit line is bonded to each 36x36$\mu$m absorber pixel. This unique bonding structure gives the readout an effective pitch of 36$\mu$m. We find the increased pitch between readout bonds significantly reduces the interpixel capacitance of the CMOS detector reported by Bongiorno et al. 2010\cite{2010SPIE.7742E..18B} and Kenter et al. 2005\cite{2005SPIE.5898..479K}. 
\end{abstract}


\section{INTRODUCTION}
\label{sec:intro} 

Current missions, such as Chandra and XMM Newton, focus on observing phenomena such as the black hole in the center of our galaxy and supernova remnants. In the future, x-ray astronomy will contribute to answering some of the most intriguing questions in the field of astronomy and astrophysics. For example dark energy can be explored by utilizing observations of hot galaxy clusters, the Warm-Hot Intergalactic  Medium (WHIM) can be explored more deeply by observing WHIM induced spectral lines in the otherwise featureless spectra of blazars, and the birth of stars and cosmic structure can be explored by discovering new x-ray bursts at early times in the universe (redshift$>$7).  New observatories are being proposed such as SMART-X, AXSIO, and ATHENA which will have more collecting area and/or enhanced spectral resolution to address the above topics.  These high rate observatories are in need of a better x-ray detector that can surpass CCD's with a faster readout that can keep up with the amount of x-rays that will be incident on the focal plane.  

Hybrid CMOS detectors can enable the next generation of observatories.  Their pixel readout architecture allows for individual pixels to be read out to constrain the source of interest before too many photons saturate a pixel.  Hybrid CMOS detectors also offer low power usage and are less susceptible to radiation damage than CCD's.  The general advantages of Hybrid CMOS detectors are described in Falcone et al. 2012 \cite{2012abe} (these proceedings).     

However, Interpixel Capacitance(IPC) has been a problem in Hybrid CMOS detectors. IPC causes charge spreading and makes single pixel events into cross events (i.e. events with significant charge in the four surrounding pixels above, below, left, and right of the central pixel).  This unwanted charge spreading means that more pixels must be read out to accurately determine the charge in the event and the energy resolution will be degraded.   The charge spreading also makes it difficult to capture all of the charge in the x-ray event, further degrading the energy resolution. 

In this paper, we show that for larger pixel sizes the IPC problem is significantly reduced.  We discuss the IPC characterization of  a Teledyne Hawaii Type H2RG Hybrid CMOS x-ray detector with an effective pitch of 36$\mu$m and compare it to three Teledyne Hawaii Type H1RG Hybrid CMOS x-ray detectors with 18$\mu$m pitch.  We first begin with a background on IPC and its effects on detector characteristics, then talk about our setup and measurement technique and finally discuss the results and implications.    

\section{Interpixel Capacitance} 
Interpixel Capacitance(IPC) is caused by unintended, parasitic capacitances between adjacent pixels.  Figure 1 illustrates a circuit diagram showing both the capacitance between each pixel and its readout node, $C_{0}$, and the IPC, $C_{c}$.  
The interpixel capacitance of a detector can make single pixel events appear as poorly constrained cross events.  When a single pixel event is detected the voltage signal will be spread between all four adjoining pixels and into the four corner pixels as well.  An x-ray event with significant IPC from one of our H1RG detectors can be seen in Figure 2.

\begin{figure}[htbp]
  \begin{minipage}[b]{0.5\linewidth}
    \centering
    \includegraphics[width=\linewidth]{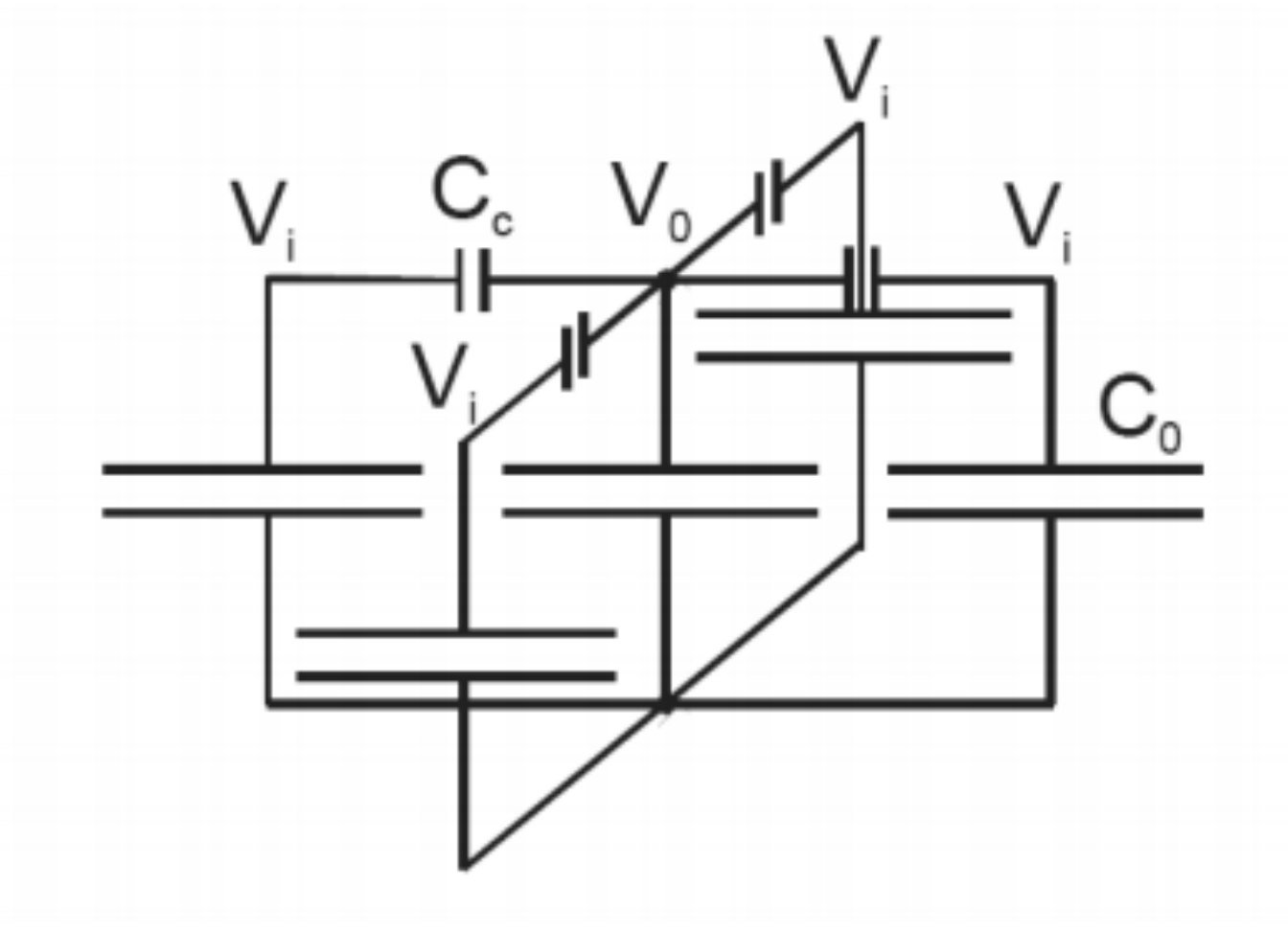}
    \caption{Schematic of inter pixel capacitance.  $V_{0}$ is the center pixel, $C_{c}$ is the interpixel capacitance, $C_{0}$ is the expected capacitance between the pixel and readout mode and $V_{i}$ are the neighboring pixels.  Figure is from Finger et al 2006\cite{2006SPIE.6276E..13F}}
    \label{fig:chapter001_dist_001}
  \end{minipage}
  \hspace{0.5cm}
  \begin{minipage}[b]{0.45\linewidth}
    \centering
    \includegraphics[width=\linewidth]{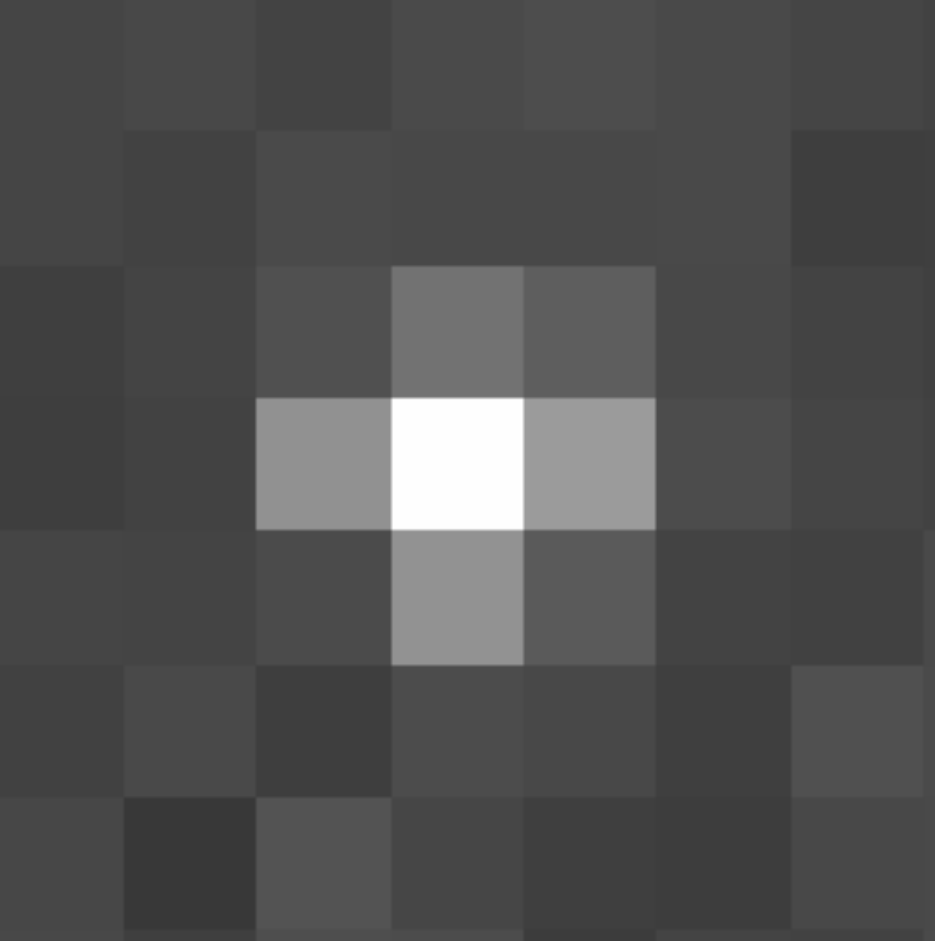}
    \caption{An x-ray event with significant IPC from a H1RG detector.}
    \label{fig:chapter001_reward_001}
  \end{minipage}
\end{figure}


In order to capture most of the charge in Figure 2, one must read out at least five pixels instead of one.  This charge spreading adds additional read noise, dark current and pixel-to-pixel gain variation which degrades the energy resolution of the detector.  The IPC also makes it difficult to capture all of the charge in the x-ray event.  A portion of the charge from the x-ray may spread outside the eight surrounding pixels and therefore not be counted in the total charge measurement.  

Due to the degrading effect of IPC on x-ray events, it is essential that next generation CMOS detectors have low IPC in order to capitalize on their other advantages over CCD's for future x-ray missions.

\section{Hybrid CMOS}
The detectors that we have been testing have been Teledyne Hawaii Type Hybrid CMOS detectors.  A schematic cross section of a Hybrid CMOS detector can be seen in Figure 3.  
\begin{figure}[h!]
\vspace{1cm}
\center
\includegraphics[width=8cm,height=4cm,scale=.75]{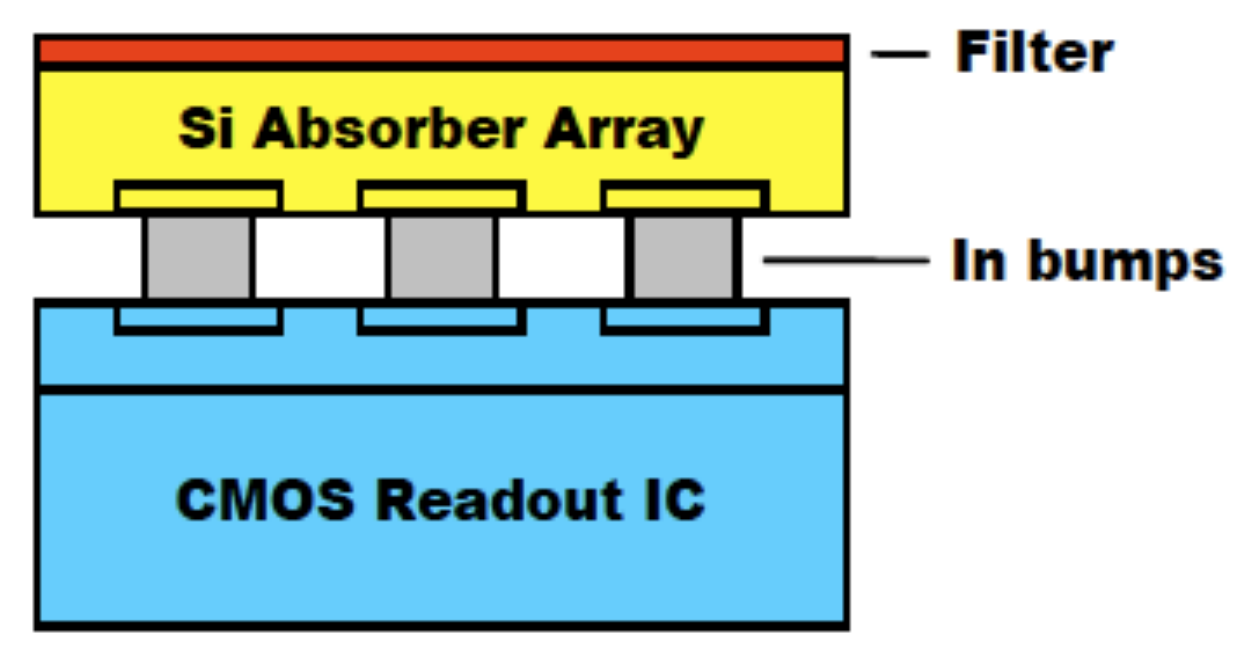}
\caption{Schematic cross-section of a hybrid CMOS detector showing an aluminum filter, absorber array, indium bumps and CMOS multiplexer  }
\end{figure}

A hybrid CMOS detector consists of two components; the absorber array which is responsible for photon-to-charge conversion
via photoelectric absorption and the readout multiplexer which functions as a charge-to-voltage converter and signal processor.  The two components are precisely aligned and connected together at every pixel through indium bonds.  This separation allows the components to be optimized independently and then connected.  The silicon absorber focuses on high quantum efficiency while the multiplexers feature high speed readout.  


The H2RG detector discussed in this paper has similar structure to the H1RG CMOS devices previously tested in that the read out array pitch is 18$\mu$m.  However, in this H2RG the absorber array is 36$\mu$m, twice the size of the absorber array in an H1RG detector, and only one readout multiplexer line is bump bonded to each absorber pixel thus leading to an effective pitch of 36$\mu$m.  This is a detector format that has been proposed for the JANUS mission \cite{2010SPIE.7732E.140F}.  The reason for having twice as much space between bonds is to lower the interpixel capacitance between the pixels. The larger gap between each pixel reduces the capacitance between neighboring pixels.    For this paper, we will discuss the measurement of IPC for the H2RG detector and compare it to the measurement of three H1RG detectors where every readout multiplexer is bump bonded to the absorber, which gives the H1RG detectors an 18$\mu$m pitch.  Table 1 summarizes the four detectors pixel pitch size.  
\begin{table}[h!]
\centering
\vspace{.2cm}
\begin{tabular}{c|c}
Detector&Pixel Pitch Size \\
\hline
H1RG-125&18$\mu$m\\
H1RG-161&18$\mu$m\\
H1RG-167&18$\mu$m\\
H2RG-122&36$\mu$m\\
\end{tabular}
\vspace{0.3cm}
\caption{The four detectors tested and their pixel pitch size.}
\end{table}


\section{Experiment Setup}
In order to test our hybrid CMOS detectors we had to first cool each detector and put it under vacuum.   For these tasks we used our ``cube" setup (also known as Testcam 2) pictured in Figure 4.

\begin{figure}[h!]
\center
\includegraphics[width=8cm,height=6cm,scale=.75]{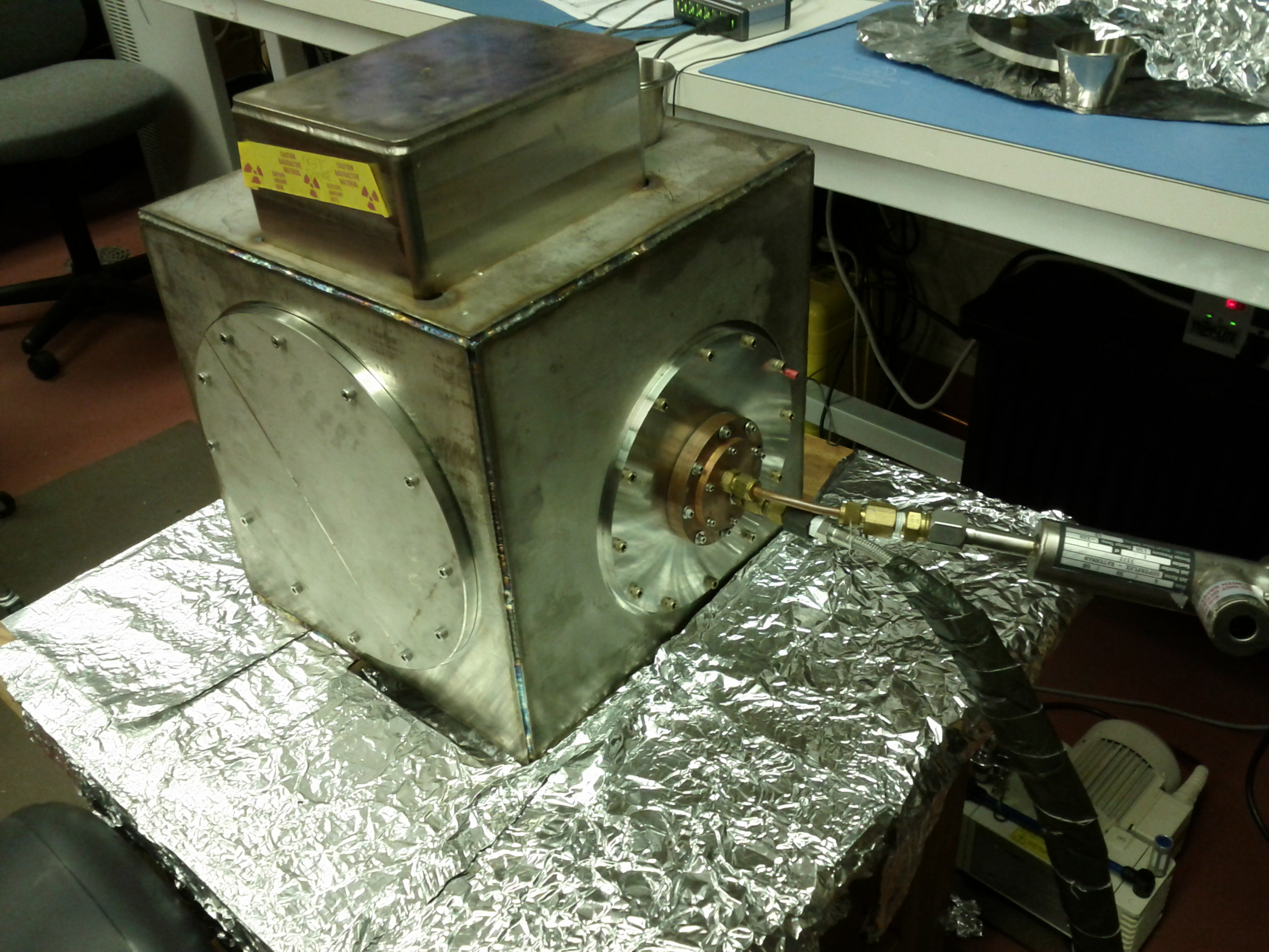}
\caption{Cube setup used to expose the H1RG and H2RG detectors to x-rays.  }
\end{figure}
Before taking any exposures, we pump the cube chamber down to $\sim10^{-6}$ millibar and cool the detector using liquid nitrogen to 150 Kelvin.  The detector temperature is controlled by a Labview VI that uses a Proportional Integrative Derivative (PID) algorithm that can regulate the temperature to within +/- 0.2K.  In this setup, the detector is run with a 15 volt bias voltage, which leads to minimal charge spreading in a 36$\mu$m pixel.  We then exposed the detector to an $^{55}$Fe x-ray radiation source (which produces Mn K$\alpha$ and K$\beta$ lines).  The cube contains a room temperature SIDECAR$^{\rm TM}$ ASIC board that generates the CMOS clock and bias signals and digitizes the output signals.  The Teledyne IDE controls the SIDECAR's registers and Teledyne's IDL JAC code controls the exposures.  We then take multiple datasets of hundreds of images which are then pseudo-CDS subtracted via software.  We then use event detection on the subtracted images to produce a spectrum.  

\subsection{X-ray Source}

We used a radioactive $^{55}$Fe source to produce x-rays. In this case, the capture of a K-shell electron by a proton results in a neutron.  The electrons then cascade down to lower energy levels emitting manganese K$\alpha$ (5.9 keV) and K$\beta$ (6.5 keV) lines.  We observe high count-rates in these x-ray lines due to the sources high activity and close proximity to the detector.
  

We then measured the IPC of the detectors by using the $^{55}$Fe x-ray source and the cube setup at a temperature of 150K to suppress the dark current.  For each detector, we took 2 datasets  of 100 images each with approximately a 5.28 second exposure time which amounted to  $\sim$100,000 x-ray events.   


\section{Measuring Interpixel Capacitance} 
As mentioned in Section 2 and shown in Figure 2, IPC makes ``cross-like" events with charge spreading up, down, and to the sides.  Therefore, we measured the IPC of the CMOS detectors by finding the most symmetrical x-ray events in the $^{55}$Fe source data at 150K.   Events we did not want to include were ``split" events.  ``Split" events occur when an x-ray charge is split between two or more pixels due to the position on the pixel where the x-ray is detected.  If the x-ray charge cloud diffuses into two or more pixels as it reaches the base of the pixel, the charge will be split between them.  We did not want to include these events because they would skew our IPC measurement since the spreading of the charge in ``split" events is not due to IPC.  Charge splitting also cannot account for cross shaped events.  Due to our high bias voltage of 15 volts, the full width half max of charge diffusion remains less than half of our pixel pitch, as seen in Bai et al 2008 \cite{2008SPIE.7021E...1B}.  Therefore, by selecting only symmetric cross-like events, we ensure that charge splitting does not contribute any significant charge to our selected events.  

We found these symmetric events by first finding high energy x-ray events containing charge above a ``primary threshold" in a single pixel where its value is larger than the surrounding eight pixels. The events must then satisfy the following criteria.  We first determined whether an event was a split event by using a ``secondary threshold".  This criteria looks at the surrounding eight pixels of the x-ray event and if one pixel is measured to be above the threshold, measured in DN, it will grade it as a split event.  We next used a region cut to avoid using x-ray events near the edge of the detector where measured gain variation is present. 
We then only used x-ray events where the secondary pixel in the x-ray event was 20\% or below the value of the primary center pixel.  We also measured the total charge of all nine pixels in the event to verify the event was from an $^{55}$Fe x-ray.  We found the mean of the manganese K$\alpha$ and K$\beta$ peak and went out 4$\sigma$ below the K$\alpha$ peak and 4$\sigma$ above the K$\beta$ peak to ensure we were getting all of the symmetrical $^{55}$Fe x-ray events. 


Since some of the above cuts will include split events in our calculation, we next used a standard deviation method to find the most symmetrical events.  We first measured the standard deviation of the four adjacent pixels of the event.  If the measured standard deviation of the charge of the four adjacent pixels was less than the 1 sigma read noise floor, we counted it as a single pixel event, which was suitable for measuring IPC.  We used this algorithm to ensure we were using single pixel events with surrounding pixels solely affected by IPC or noise, as opposed to the charge spreading mentioned above. We then averaged and normalized all nine pixels of the IPC events to get a normalized kernel for each detector. We did not include any events whose adjacent four pixels standard deviations were above the read noise since we wanted to solely include single pixel events.   We therefore used the read noise as a cutoff for IPC events which gave us an upper limit on the IPC in the detector. 

\subsection{H1RG Detectors}
We used the method described above to test the H1RG detectors with an 18$\mu$m pitch.  Two of the detectors, H1RG-125 and H1RG-161, had their IPC previously characterized using a different method.  Bongiorno et al. \cite{2010SPIE.7742E..18B} used a method in which 
they selected IPC events where the second brightest pixel in the event is between 0\% and 15\% of the value of the brightest pixel.  Their goal was to select the most symmetric x-ray events using this method.  The IPC kernels of H1RG-125 and H1RG-161 using the second brightest pixel method from Bongiorno et al. \cite{2010SPIE.7742E..18B} and our results using the standard deviation method can be seen in Tables 2 and 3.

\begin{table}[ht]
\begin{minipage}[b]{0.5\linewidth}\centering
\begin{tabular}{|c|c|c|}
\multicolumn{3}{c}{Second Brightest Pixel}\\
\hline
0.013&0.047& 0.012 \\
\hline
0.076&0.701&0.068\\
\hline
0.014&0.049&0.013\\
\hline
\end{tabular}
\end{minipage}
\hspace{0.1cm}
\begin{minipage}[b]{0.0\linewidth}
\centering
\begin{tabular}{|c|c|c|}
\multicolumn{3}{c}{Standard Deviation}\\
\hline
0.019 $\pm$ 0.009&0.070 $\pm$ 0.027& 0.020 $\pm$ 0.011 \\
\hline
0.070 $\pm$ 0.024&0.630 $\pm$ 0.14&0.074 $\pm$ 0.024\\
\hline
0.024 $\pm$ 0.019&0.070 $\pm$ 0.024&0.022 $\pm$ 0.013\\
\hline
\end{tabular}
\end{minipage}
\vspace{0.3cm}
\caption{IPC Kernel for H1RG-125 using the second brightest pixel method\cite{2010SPIE.7742E..18B} and the standard deviation method.}
\end{table}

\begin{table}[ht]
\begin{minipage}[b]{0.5\linewidth}\centering
\begin{tabular}{|c|c|c|}
\multicolumn{3}{c}{Second Brightest Pixel}\\
\hline
0.012&0.046& 0.012 \\
\hline
0.072&0.712&0.071\\
\hline
0.012&0.047&0.012\\
\hline
\end{tabular}
\end{minipage}
\hspace{0.1cm}
\begin{minipage}[b]{0.0\linewidth}
\centering
\begin{tabular}{|c|c|c|}
\multicolumn{3}{c}{Standard Deviation}\\
\hline
0.0088 $\pm$ 0.010&0.063 $\pm$ 0.02& 0.0094 $\pm$ 0.009 \\
\hline
0.067 $\pm$ 0.022&0.71 $\pm$ 0.14 &0.061 $\pm$ 0.02\\
\hline
0.011 $\pm$ 0.013&0.064 $\pm$ 0.022&0.0082 $\pm$ 0.009\\
\hline
\end{tabular}
\end{minipage}
\vspace{0.3cm}
\caption{IPC Kernel for H1RG-161 using the second brightest pixel method\cite{2010SPIE.7742E..18B} and the standard deviation method.}
\end{table}

Overall our results using the standard deviation method are consistent with the results from Bongiorno et al.\cite{2010SPIE.7742E..18B}.  Both the H1RG-125 and H1RG-161 retain 64\% to 70\% of the x-ray signal in the center pixel while $\sim$6\%-7\% is spread into one of the four adjoining pixels.  

We next characterized the IPC of H1RG-167 for the first time using the standard deviation method.  The IPC kernel can be seen in Table 4.  
\begin{table}[h!]
\centering
\vspace{.2cm}
\begin{tabular}{|c|c|c|}
\hline
0.017 $\pm$ 0.0037&0.077$\pm$ 0.010& 0.018 $\pm$ 0.0037 \\
\hline
0.082 $\pm$ 0.009&0.61 $\pm$ 0.049&0.081 $\pm$ 0.009\\
\hline
0.017 $\pm$ 0.004&0.080 $\pm$ 0.014&0.018 $\pm$ 0.004\\
\hline
\end{tabular}
\vspace{0.3cm}
\caption{IPC Kernel for H1RG-167 using the standard deviation method}
\end{table}
Overall the IPC of the the three H1RG detectors is approximately 6-8\%  per each adjacent pixel and approximately 1\% per each corner pixel.  

\subsection{H2RG Detector}
We next tested the H2RG-122 detector with the 36$\mu$m effective pitch.  We used the standard deviation method to get the IPC kernel seen in Table 5.  
\begin{table}[h!]
\centering
\vspace{.2cm}
\begin{tabular}{|c|c|c|}
\hline
0.007 $\pm$ 0.01&0.016 $\pm$ 0.009& 0.007 $\pm$ 0.011 \\
\hline
0.017 $\pm$ 0.009&0.905 $\pm$ 0.11 &0.017 $\pm$ 0.010\\
\hline
0.007 $\pm$ 0.01&0.017 $\pm$ 0.009&0.007 $\pm$ 0.011\\
\hline
\end{tabular}
\vspace{0.3cm}
\caption{IPC Kernel for H2RG-122 using the standard deviation method}
\end{table}
The IPC for the H2RG-122 detector is significantly less than the H1RG detectors.  In the H2RG detector, 90.5\% is retained in the center pixel and approximately 1.7\% is spread into each of the four adjacent pixels and 0.7\% spread into the corner pixels.  This result shows the 36$\mu$m effective readout pitch significantly reduces the IPC in the detector.  

We characterized the uncertainty in the IPC values by individually calculating the standard deviation for each pixel in the nine pixel events, across all events in the sample. We then propagated the errors through the normalization of the average IPC kernel.  We believe that this calculation of the error accounts for error due to both read noise as well as the unintentional inclusion of split events in the sample.

\section{Conclusion}
Overall we have developed a method for measuring the Interpixel Capacitance in a Hybrid CMOS detector and have measured the IPC of four hybrid CMOS detectors detectors including three H1RG's with 18$\mu$m pitch and one H2RG with a 36$\mu$m effective pitch.  We found the H1RG detectors with an 18$\mu$m pitch had IPC of approximately 6-8\% per adjacent pixel and the H2RG had an IPC of approximately 1.7\% per each adjacent pixel.  This result shows that the unique bump bonding pattern of the H2RG significantly reduces the IPC seen in Hybrid CMOS detectors.  This solution to the IPC problem will allow less charge spreading, which will constrain the x-ray charge into fewer pixels and decrease the multi-pixel accumulated read noise and loss of charge in x-ray events.  These factors will combine to improve the energy resolution in future Hybrid CMOS detectors.  The readout architecture offers a solution to IPC in CMOS detectors as they continue to be a viable option for future x-ray missions.  We are currently working with Teledyne to modify the ROIC architecture in a way that will allow the x-ray hybrid CMOS detectors to have unmeasurable IPC for small pixel ($<$ 18$\mu$m) devices.       

\section{Acknowledgements} 
We gratefully acknowledge Teledyne Imaging Systems, particularly James Beletic and Yibin Bai, for providing useful comments and for loaning us the modified H2RG detector.  This work was supported by NASA grants NNG05WC10G, NNX08AI64G, and NNX11AF98G.

\bibliography{SPIEipcpaper}   
\bibliographystyle{spiebib}   

\end{document}